\documentclass[aps,prr,reprint,superscriptaddress]{revtex4-2}

\usepackage{amsmath}
\usepackage{amssymb}
\usepackage{hyperref}
\usepackage[capitalize]{cleveref}
\usepackage{physics}
\usepackage{svg}
\usepackage{orcidlink}
\usepackage[acronym]{glossaries-extra}
\usepackage[caption=false]{subfig}
\usepackage{csquotes}

\glsdisablehyper
\setabbreviationstyle[acronym]{long-short}

\newacronym{hhg}{HHG}{high-order-harmonic generation}
\newacronym{td-dft}{TD-DFT}{time-dependent density functional theory}
\newacronym{dft}{DFT}{density functional theory}
\newacronym{tdlwf}{TDLWF}{time-dependent localized Wannier function}
\newacronym{mlwf}{MLWF}{maximally localized Wannier function}
\newacronym{hgh}{HGH}{Hartwigsen-Goedecker-Hutter}
\newacronym{lda}{LDA}{local density approximation}
\newacronym{hbn}{h-BN}{hexagonal boron nitride}
\newacronym{dfg}{DFG}{Deutsche Forschungsgemeinschaft}

\begin{document}
    \title{Efficient time dependent Wannier functions for ultrafast dynamics}
    
    \author{Cristian M. Le\orcidlink{0000-0001-5122-8521}}
    \email{cristian.le@mpsd.mpg.de}
    \affiliation{Max Planck Institute for the Structure and Dynamics of Matter and Center for Free Electron Laser Science, 22761 Hamburg, Germany}

    \author{Hannes~H\"ubener\orcidlink{0000-0003-0105-1427}}
    \email{hannes.huebener@mpsd.mpg.de}
    \affiliation{Max Planck Institute for the Structure and Dynamics of Matter and Center for Free Electron Laser Science, 22761 Hamburg, Germany}

    \author{Ofer Neufeld\orcidlink{0000-0002-5477-2108}}
    \email{ofer.neufeld@gmail.com}
    \affiliation{Schulich Faculty of Chemistry, Technion - Israel Institute of Technology, Haifa, Israel}
    \affiliation{Max Planck Institute for the Structure and Dynamics of Matter and Center for Free Electron Laser Science, 22761 Hamburg, Germany}

    \author{Angel~Rubio\orcidlink{0000-0003-2060-3151}}    
    \email{angel.rubio@mpsd.mpg.de}
    \affiliation{Max Planck Institute for the Structure and Dynamics of Matter and Center for Free Electron Laser Science, 22761 Hamburg, Germany}
    \affiliation{Center for Computational Quantum Physics (CCQ), The Flatiron Institute, 162 Fifth avenue, New York NY 10010.}
    

    \begin{abstract}
        Time-dependent Wannier functions were initially proposed as a means for calculating the polarization current in
        crystals driven by external fields.
        In this work, we present a simple gauge where Wannier states are defined based on the maximally localized
        functions at the initial time, and are propagated using the time-dependent Bloch states obtained from
        established first-principles calculations, avoiding the costly Wannierization at ech time step.
        We show that this basis efficiently describes the time-dependent polarization of the laser driven system through
        the analysis of the motion of Wannier centers.
        We use this technique to analyze highly nonlinear and non-perturbative responses such as high harmonic
        generation in solids, using the \glsfmtlong{hbn} as an illustrative example, and we show how it provides an
        intuitive picture for the physical mechanisms.
    \end{abstract}
    \maketitle

    \section{Introduction}

    Wannier functions\cite{Wannier_1937,Marzari_2012} form a single-particle orthornormal basis in solid-state crystals
    that are widely used as an alternative to Bloch orbitals.
    Especially, \glspl{mlwf} have proven to be of great practical use, because they give chemical intuition to
    first-principles numerical results of static and thermodynamic properties\cite{Marzari_2012}.
    In this representation one can visualize properties typically studied in molecular systems, such as hybridization,
    covalency and ionicity, in solid-state crystals\cite{AbuFarsakh_2007, Evarestov_2003, Cangiani_2004}, and provide
    the basis for the implementation of the modern theory of polarization\cite{KingSmith_1993,Resta_1994,Resta_2007}.
    Wannier functions also play an important role in the formulation and evaluation of response properties, because of
    their close relation to polarization\cite{Vanderbilt_2006}, Born effective charges\cite{Sony_2008}, Berry phase and
    curvature\cite{Tsirkin_2021} etc., making them particularly useful in describing various topology related
    properties.
    The Wannier functions have been expanded to perturbation theory\cite{Lhim_2021}, Floquet theory\cite{Nakagawa_2020},
    and time propagation\cite{Souza_2004,Yost_2019} in order to obtain localized basis representations in each of these
    domains.
    They have also been used as interpolators to extract electron-phonon coupling\cite{Giustino_2007, Noffsinger_2010,
    Ponc_2016, Zhou_2021b}.

    \Glspl{mlwf} are widely used as a convenient and well-defined orbital basis to derive tight-binding models of
    materials from first principles.
    These models are especially useful in analyzing complex materials such as highly correlated systems\cite{Ku_2002},
    because they allow the reduction of the correlated problem to a set of few active atomic orbitals per site, for
    instance through methods like constrained RPA\cite{Aryasetiawan_2011}.
    To describe dynamical properties in such systems one normally propagates these lattice models built on top
    of a static \glspl{mlwf} basis set.
    This requires in principle the wannierization of all unoccupied states into which the system evolves, relying on
    prior intuition of the dynamic processes involved and often entailing increased computational cost.

    Another proposed approach is to make the Wannier functions themselves time-dependent and calculate the observables
    directly in this basis\cite{Souza_2004}.
    Despite the early recognition of this application, Wannier function propagation has not been systematically pursued
    as a practical and efficient computational strategy and the precise definition has not yet been explored.
    The one notable case is an approach where the wannierization is done with the constraint to maximally localize
    the states at each time-step of the out-of-equilibrium driven system\cite{Yost_2019, Zhou_2021}.
    This approach showed good agreements between the electronic polarization and experimental absorption
    spectra\cite{Yost_2019}, and it retained the intuitive chemical representation in that process.
    However, besides the high computational cost associated with maximally localizing the Wannier functions at
    each time-step, there is also a conceptual issue in that this basis set can have discontinuities in time as the
    global minima of the \glspl{mlwf} changes, leading to discontinuous derived observables\cite{Yost_2019}.

    The most direct application of time-dependent Wannier functions is to study the dynamics of the dielectric
    polarizations and its current, particularly beyond the regime of response theory, for instance
    \gls{hhg} in solids which is not accessible via perturbation theory\cite{Ghimire_2018, Ghimire_2014, Yue_2022}.
    With the advances in ultrafast lasers, \gls{hhg} in solid-state materials have been studied both experimentally
    and computationally.
    One common calculation method is using tight-binding models\cite{Mrudul_2021, Rana_2022, Bauer_2018, Jr_2019,
    Silva_2019b, Baykusheva_2021, Chacn_2020, Lv_2021, Luu_2018}, but in order to fully capture the dynamics,
    particularly the interband dynamics, these models need to be expanded to include the unoccupied
    bands\cite{Ikemachi_2017}, increasing the computational cost.
    \Glspl{mlwf} have been shown to be especially useful for analyzing \gls{hhg}, but the basis set here is static
    corresponding to the ground state system\cite{Silva_2019, Brown_2024, Osika_2017, Catoire_2018, Parks_2020,
    Murakami_2018}, unlike other useful basis sets which evolve with the Hamiltonian, like Houston\cite{Krieger_1986,
    Sato_2014, Sato_2018, Byun_2021, Yue_2022} or
    Floquet-Bloch states\cite{Faisal_1997, Jin_2019, Wu_2015, Ikeda_2018, Galler_2023, Neufeld_2022, Dimitrovski_2017}.
    On the other hand, first-principles methods, like the real-time \gls{td-dft}\cite{Marques_2012}, incorporate all of
    these dynamics within the occupied orbitals only\cite{Neufeld_2023, Neufeld_2022b, Zhang_2024, Neufeld_2023b,
    TancogneDejean_2022, TancogneDejean_2017, TancogneDejean_2018}.
    Here we show that the time evolution of the Wannier functions are able to incorporate all of these dynamics within a
    far smaller basis by combining these methods.

    Here we propose and numerically study a definition of the time-dependent Wannier functions,
    where the wannierization step is done only once on the ground state (i.e at $t=0$), and the time-dependent Wannier
    states are determined by the time-dependent Bloch states, which we take from established first-principles
    calculations.
    We use this definition to study \gls{hhg} in both non-interacting model system and ab-initio in \gls{hbn}.
    We show that with this definition, the current can be described purely from the motion of Wannier centers in a wide
    range of laser driving profiles, providing an intuitive analogue for the dipole acceleration in finite systems but
    applied to periodic solids.

    \section{Time dependent Wannier function definitions} \label{sec:td-wannier-func}

    Here, we consider time-dependent Wannier functions of the following form
    \begin{gather}
        \Phi_{n\vb{R}}(\vb{r},t)=\sum_{i}\int_{BZ}U^{n\vb{R}}_{i\vb{k}}f_{i\vb{k}}(t)\Psi_{i\vb{k}}(\vb{r},t)\dd[3]{\vb{k}}, \label{eq:td_wannier}
    \end{gather}
    where $\Phi$ are the time-dependent Wannier functions, $\Psi$ are the time-dependent Bloch states, $U$ is the
    Wannier transformation matrix, which is calculated only once at the ground state to give the \gls{mlwf}, and $f$ is
    a time-dependent phase compensation function that we introduce in order to retain the localized nature of the
    Wannier functions.
    Note, that one could also re-compute $U$ over time, but this would not be numerically efficient or guarantee a
    continuous evolution of the time-dependent Wannier states.
    Instead, we here express the time-dependence of the Wannier functions via the phase factor $f$ and time-dependent
    Bloch states $\Psi$ only, which does not result in maximally localized functions, but it does guarantee continuous
    evolution while still maintaining a high degree of localization.
    We explore three such phase compensation functions
    \begin{gather}
        f^{\text{none}}_{i\vb{k}}(t)=1\quad\forall i,\vb{k},t, \label{eq:f_none}\\
        f^{\text{static}}_{i\vb{k}}(t) = e^{i\epsilon^0_{i\vb{k}}t}, \label{eq:f_static}\\
        f^{\text{int}}_{i\vb{k}}(t) = e^{i\int_0^t\epsilon_{i\vb{k}}(\tau)\dd{\tau}}, \label{eq:f_int}
    \end{gather}
    where we either do not perform any phase compensation (\cref{eq:f_none}), compensate for the static band energy
    $\epsilon^0$ (\cref{eq:f_static}), or compensate for the instantaneous energy expectation value (\cref{eq:f_int})
    \begin{equation}
        \epsilon_{i\vb{k}}(t) = \ev{\hat{H}(t)}{\Psi_{i\vb{k}}(t)}. \label{eq:epsilon_i}
    \end{equation}
    The reasoning behind these choices is to explore the scope and importance of such phase compensation.
    The $f^{\text{static}}$ is the trivial initial state, or field free, evolution, while $f^{\text{int}}$ takes the
    driving field into account, via the expectation value of the time-dependent Hamiltonian $\ev*{\hat{H}(t)}$, as well
    as changes in the character of the Bloch states during time-evolution.
    To analyze the differences of these definitions, we look at the evolution of the Wannier spread $\Omega(t)$:
    \begin{gather}
        \Omega_n(t) = \int_V\vb{r}^2\abs{\Phi_{n\vb{R}}(\vb{r},t)}^2\dd[3]{\vb{r}} - \abs{\vb{r}_{n\vb{R}}(t)}^2,\\
        \vb{r}_{n\vb{R}}(t) = \int_V\vb{r}\abs{\Phi_{n\vb{R}}(\vb{r},t)}^2\dd[3]{\vb{r}},
    \end{gather}
    where we denote here $\vb{r}_{n\vb{R}}(t)$ to be the time-dependent Wannier center.

\begin{figure*}
        \centering
        \subfloat{\label{fig:bandstructure}\includegraphics[scale=0.7]{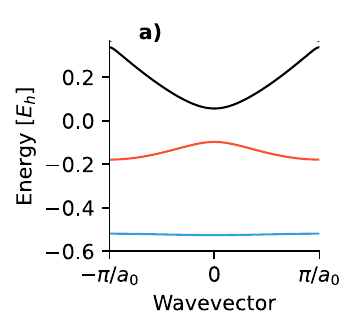}}
        \subfloat{\label{fig:mathieu_spread1}\includegraphics[scale=0.7]{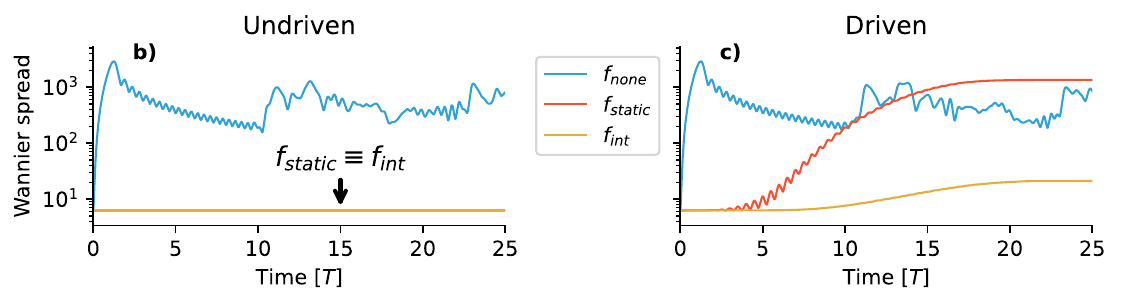}}
        \subfloat{\label{fig:mathieu_spread2}}
        \caption{Non-interacting 1D electron system under Mathieu potential.
        \protect\subref{fig:bandstructure} Ground state bandstructure, \protect\subref{fig:mathieu_spread1} and
        \protect\subref{fig:mathieu_spread2} Wannier spread evolution without and with laser perturbation respectively.
        The band gap in \protect\subref{fig:mathieu_spread1} measures to $\approx 0.154(E_h)$.}
    \end{figure*}

    We investigate the different time-dependent Wannier functions first by considering a non-interacting 1D
    electron system under a Mathieu potential with or without an external laser interaction $\vb{A}(t)$
    \begin{gather}
        \hat{H}(t) = \frac{1}{2m}(\hat{\vb{p}}-e\vb{A}(t))^2 + \hat{V},\\
        V(x) = -V_0(1+\cos(2\pi x/a)).
    \end{gather}
    Here we choose some well studied parameters for the Mathieu potential $V_0=0.37(E_h)$, $a_0=8(a_0)$,
    resulting in the static band structure shown in \cref{fig:bandstructure}.
    In this model we take the first two bands to be fully occupied, and for this section we focus on the dynamics
    of these valence band and the evolution of their corresponding Wannier function.

    We first consider the stationary ground-state case
    \begin{equation}
        \vb{A}(t)=0 \quad \forall t,
    \end{equation}
    and let both the system and \cref{eq:td_wannier} evolve coherently in time.
    By definition, all observables are time-independent and in the ground state, as the system evolves with only a pure
    global phase which cancels out when evaluating observables.
    But the Wannier function basis from \cref{eq:td_wannier} is still time-dependent as it is obtained by summing
    up Bloch states that coherently interfere due to the different phase propagations of the band eigenvalues of the Bloch
    states (\cref{fig:bandstructure}).
    This effect is evident in \cref{fig:mathieu_spread1} when evaluating the Wannier spread $\Omega$ without external
    perturbation.
    Wannier functions $f^{\text{none}}$ that do not do phase compensation (blue line) loose localisation very
    fast, with a timescale matching roughly the band energy bandwidth.
    This trivial dephasing effect is exactly compensated by the other two definitions (red and yellow lines) that show a
    spread that is independent of time.
    In \cref{fig:mathieu_spread1}, the red and yellow lines are on top of each other because by definition they are
    equivalent ($f^{\text{static}}\equiv f^{\text{int}}$) when there is no time-dependent perturbation.
    This behavior is the main motivation for the introduction of the phase compensation term $f$ in our definition of
    the time-dependent Wannier functions (\cref{eq:td_wannier}).

    Next, we consider the strongly driven non-perturbative case with a super-sine laser envelope\cite{Neufeld_2019}
    \begin{equation}
        \label{eq:A_xt}
        A_x(t)=A_0\sin(\frac{\pi t}{n_{\text{cyc}}T})^{\frac{\pi}{\sigma}\abs{\frac{t}{n_{\text{cyc}}T}-\frac{1}{2}}}\sin(\frac{2\pi t}{T}),
    \end{equation}
    where we use the standard notations for the laser period $T$, and wavelength $\lambda$
    \begin{equation}
        T = \frac{\lambda}{c}.
    \end{equation}
    Here we look at the case with laser parameters: a peak laser intensity of $I=10^{11} (W/cm^2)$, a wavelength of
    $\lambda=1500 (nm)$, corresponding to $\approx0.198$ of the direct band gap, and a pulse duration of
    $n_{\text{cyc}}=25$ cycles.
    The Wannier spread evolution in this case are represented by \cref{fig:mathieu_spread2}.
    The uncompensated evolution of the Wannier function ( $f^{\text{none}}$, blue line) shows no significant difference
    from the undriven case, and we can disregard it as a practical defieniton of time-dependent Wannier functions.
    Looking at the $f^{\text{static}}$ Wannier function (red line), we see that compensating for the static
    energy dephasing alone is also insufficient to retain the localized nature, as the spread evolves over time to
    become as fully delocalised as the uncompensated case.
    This is particularly evident when compared to the performance of $f^{\text{int}}$ Wannier function (yellow line),
    which looses only some locality during this strong driving process.
    Note that the spread in \cref{fig:mathieu_spread1,fig:mathieu_spread2} are shown on a logarithmic scale.
    We observe that even with the phase compensation of \cref{eq:f_int}, the Wannier functions still retain some
    dynamics, and in the next section, we show that these dynamics can carry physical significance.

    \begin{figure}
        \centering
        \includegraphics[scale=0.7]{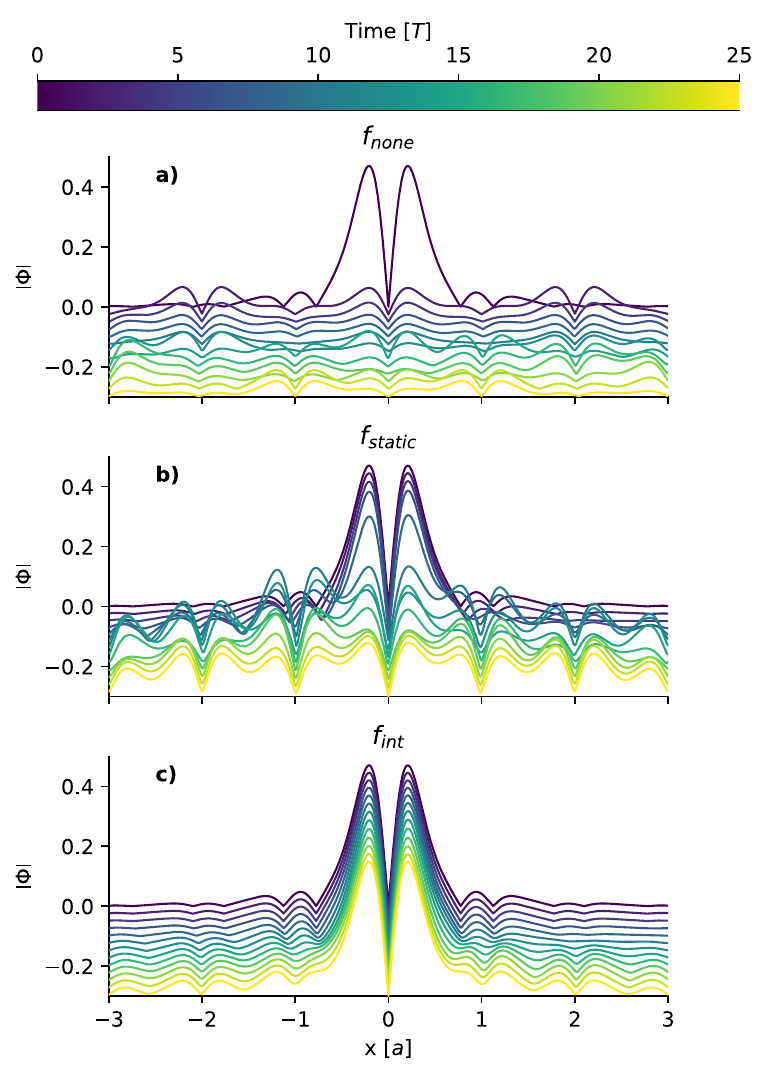}
        \caption[]{Propagation of the Wannier function envelope in time. Each evaluated time-step is color coded
        according to the colorbar and shifted downwards.}
        \label{fig:mathieu_wf}
    \end{figure}

    The model system here is simple enough that we can explore the evolution of the orbitals themselves in
    \cref{fig:mathieu_wf}, where we plot the propagation of the Wannier function envelope for the case of the strong
    driving.
    Here we get a more intuitive view of how the $f^{\text{int}}$ Wannier functions retain their localized nature.
    We can see that, although subtle, the Wannier function envelope is still evolving with the appearance of small
    oscillations in the tails of the enveloped.
    This indicates that there is some non-trivial dynamics encoded in the Wannier-center and -spread.
    The time-dependent Wannier centers allow us to generalize the theory of polarization to the time-domain, as
    discussed in the next section.

    We note that this definition of time-dependent Wannier orbitals is closely related to other commonly used
    time-dependent single particle basis sets, such as the Peierls basis and the Houston basis, which we detail in
    \cref{sec:relation-to-other-basis}.
    We especially note that the other basis do not change the shape of the single particle wave function envelope, and
    thus do not vary in their center and spread, as we have with the time-dependent Wannier functions given here.
    In \cref{sec:relation-to-other-basis} we also give a reasoning of why \cref{eq:f_int} is expected to work well for
    off-resonant excitation.
    Finally, we point out that here we do not discuss the generation of a time-dependent Wannier basis, but rather the
    construction of such states from a given time-evolution of Bloch states.
    This means that the time-dependence of the Bloch states is required to be available through an explicit
    time-propagation method and the Wannier functions discussed here can be thought of as a co-evolving complement basis
    that ideally spans the same space, but with far fewer states.

    \section{High-harmonics generation with time-dependent Wannier functions} \label{sec:results}

    In order to showcase the capability of the time-dependent Wannier functions as a basis set, we investigate the
    dynamic theory of polarization derived from the time-dependent Wannier centers\cite{Souza_2004}.
    We calculate the time-dependent Wannier function for a model non-interacting 1D system and a real-space \gls{td-dft}
    calculation on \gls{hbn}.
    The calculations here were performed using \textit{octopus}\cite{Tancogne_Dejean_2020}.

    The main observable that we investigate is the polarization current derived from the electric dipole moment:
    \begin{equation}
        \label{eq:j_wan}
        \vb{J}(t) = \dv{\vb{P}(t)}{t},
    \end{equation}
    where $\vb{J}$ is the total electronic current and $\vb{P}$ is the dynamic electronic polarization which we relate
    to the time-dependent Wannier functions\cite{Souza_2004} via
    \begin{equation}
        \label{eq:p_wan}
        \vb{P}(t) = -\frac{e}{V}\sum_{n}\vb{r}_{n\vb{R}}(t).
    \end{equation}
    We compare this current with the integrated current density calculated directly in the Bloch state
    representation\cite{Tancogne_Dejean_2020} taken to be the reference current.

    Here we analyze the \gls{hhg} yield\cite{TancogneDejean_2017,TancogneDejean_2022,Neufeld_2023,Neufeld_2023b,Zhang_2024b}
    \begin{equation}
        \xi(t) = \dv{\vb{J}(t)}{t},
    \end{equation}
    which we calculate either directly from the integrated current density evaluated in the full \gls{td-dft}
    calculation by taking the time-dependent expectation value of the current operator, or from the polarization current
    that is in turn derived from the Wannier centers.
    We are primarily investigating the spectra of the \gls{hhg} yield which we calculate from its Fourier transform
    \begin{equation}
        \xi(E) = \int e^{-iEt} \xi(t) \dd{t}.
    \end{equation}
    We apply a Gaussian filter along the time-domain before evaluating its Fourier transform in order to clean up the
    residual oscillation at the tail of the laser pulse.

    \subsection{Non-interacting 1D system} \label{sec:Mathieu}

\begin{figure*}
        \centering
        \includegraphics[scale=0.7]{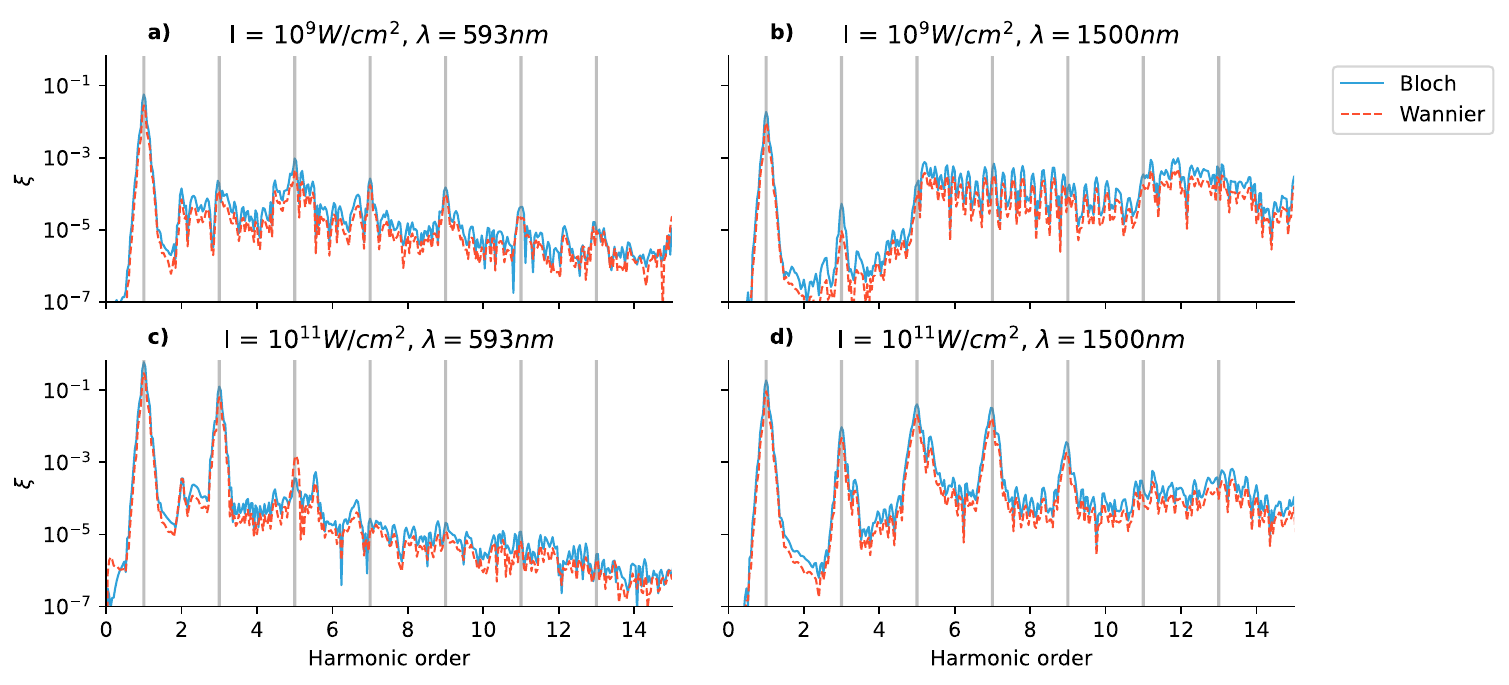}
        \caption[]{\gls{hhg} spectra of the non-interacting 1D system with varying laser interactions. The different
        lines correspond to the \gls{hhg} yield calculated from the current calculated directly from the Bloch states
        or from the Wannier centers (\cref{eq:j_wan}).}
        \label{fig:mathieu_hhg}
    \end{figure*}

    For the non-interacting 1D system we again take the Mathieu potential system that we explored in
    \cref{sec:td-wannier-func}, and we explore a wider range of laser parameters.
    The \gls{hhg} in this system has been extensively studied using various calculation
    methods\cite{Ikemachi_2017,Liu_2017}.
    In \cref{fig:mathieu_hhg} we see a good agreement of the \gls{hhg} yield derived from the Wannier functions (dashed
    red line) to the reference calculation using the Bloch states (solid blue line), across a wide range of laser
    parameters.

\begin{figure}
        \centering
        \includegraphics[scale=0.7]{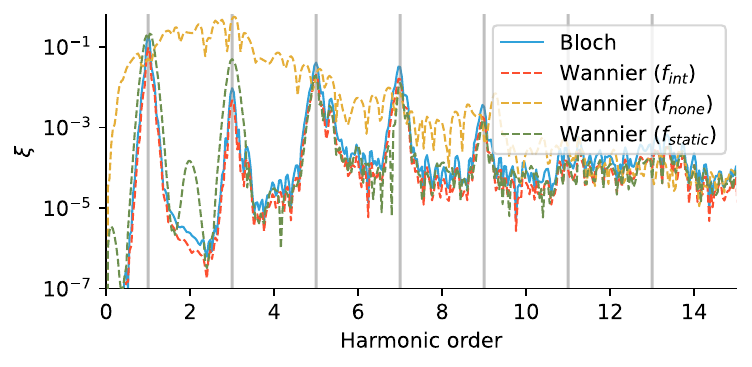}
        \caption[]{\gls{hhg} spectra of the non-interacting 1D system of the different time-dependent Wannier function
        definitions from \cref{eq:f_none,eq:f_static,eq:f_int}}
        \label{fig:mathieu_hhg_defs}
    \end{figure}

    It is worth emphasizing how good the agreement of these results is.
    We reiterate that the time-dependent Wannier function definition is not unique, so there is no guaranteed
    correlation between the current obtained from the Wannier functions through \cref{eq:p_wan,eq:j_wan}, and the
    observable current calculated from the Bloch states.
    This was clear from \cref{sec:td-wannier-func}, but it is even more evident if we compare the \gls{hhg} spectra from
    \cref{fig:mathieu_hhg_defs}, where it is clear that only the definition using \cref{eq:f_int,eq:td_wannier} gives us
    a polarization current that matches the reference observable.
    In principle, we would still be able to use the other Wannier functions as a time-dependent basis set, but in order
    to calculate an observable on such a basis, we would have to calculate the time-dependent occupation and/or expand
    the time-dependent Wannier basis set beyond the initially occupied ones in order to compensate for the difference
    in the Hilbert space.
    Here we show that time-dependent Wannier basis is an efficient basis set for capturing the
    local dynamics of the electrons in a more compact and localized representation.
    Importantly, this representation is able to capture the full range of non-linear processes in \gls{hhg}, as can be
    seen in the agreement of the whole \gls{hhg} spectra
    evaluated in \cref{fig:mathieu_hhg}.

\begin{figure}
        \centering
        \includegraphics[scale=0.7]{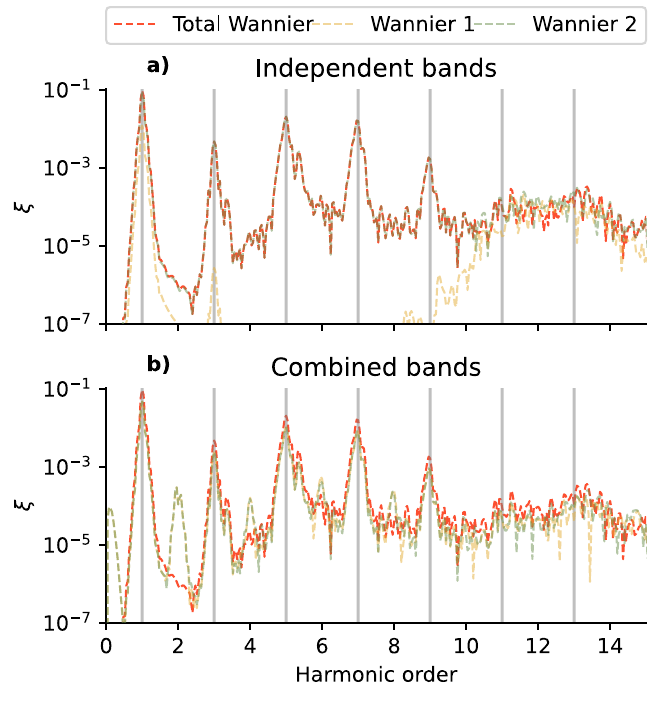}
        \subfloat{\label{fig:mathieu_hhg_wfs_independent}}
        \subfloat{\label{fig:mathieu_hhg_wfs_combined}}
        \caption[]{\gls{hhg} spectra of the non-interacting 1D system with different initial Wannier functions.}
        \label{fig:mathieu_hhg_wfs}
    \end{figure}

    One key usage of the Wannier functions is to gain chemical intuition of electronic structure and is worthwhile
    exploring how much of an intuitive picture of the dynamic processes is provided by our time-dependent Wannier
    functions.
    We now show that the time-dependent Wannier functions, indeed allow for an intuitive picture also for processes in
    the time domain and even in highly non-linear ones.
    First we Wannierize each band in \cref{fig:bandstructure} separately, i.e. without performing the sum over bands in
    \cref{eq:td_wannier}, and we propagate these states individually according to \cref{eq:f_int,eq:td_wannier}.
    We plot the contribution to the \gls{hhg} spectra of each of these Wannier functions in \cref{fig:mathieu_hhg_wfs},
    where we observe that effectively all of the \gls{hhg} is contained in the dynamics of the top-valence Wannier
    function, even as the lower band Wannier function shows an equally strong linear response.

    Next we perform the Wannierization procedure on both bands together, as done in all examples above, allowing the
    Wannier functions to mix and converge to a more localized basis set, and we propagate these states individually as
    before.
    This is more indicative to the general procedure that we might encounter when the bands are not as well separated
    that we could neatly converge the individual bands.
    What is immediately intriguing about the result shown in \cref{fig:mathieu_hhg_wfs_combined} is that the total
    \gls{hhg} remains unchanged, which can be explained by how the initial Wannier functions are related with each other
    by a unitary transformation.
    In this representation, the dynamics of the valence band is spread across both Wannier functions, and therefore we
    cannot visualize the \gls{hhg} as nicely separated contributions as before.
    However, we can still observe an important effect: the individual Wannier states propagate with even order
    harmonics  that cancel each other out due to the dynamical inversion symmetry in the model\cite{Neufeld_2019b}.
    This shows that the individual trajectories of Wannier centers of mixed band character can break the symmetry,
    however the dynamics of the whole set of Wannier centers is restoring the symmetry.
    This can be thought of in analogy to the motion of individual atoms in phonons modes, that only collectively form
    the eigenmode of the lattice.

    \subsection{Hexagonal-BN} \label{sec:hbn}

    To show the generality of this approach we now consider a realistic system in 3D, \gls{hbn} which we drive with
    similarly strong laser profiles.
    Although the system itself is a 2D lattice, and we only present the non-trivial results that are observed along the
    lattice plane, the calculations are performed in the full 3D real-space grid.
    The simulation is done with \gls{hgh} pseudo-potentials\cite{Hartwigsen_1998} and corresponding \gls{lda}
    exchange-correlation functionals, and lattice dimension of $a=4.76 (a_0)$.
    For the laser driving we take only $n_{\text{cyc}}=8$ and focus on the results with a peak laser intensity of
    $I=10^{11} (W/cm^2)$ and wavelength of $\lambda=800 (nm)$.
    The simulation was done in a real-space grid with spacing $\Delta a=0.38 (a_0)$, a $k$-space grid of
    $27\times27\times1$, and a time step of $\Delta t=0.29 (\hbar/E_h)$.
    In this simulation we only sample the time-dependent Wannier functions every 20 time steps.

    The \gls{hhg} spectra of this system has previously been investigated using first-principles methods and model calculations
    alike\cite{Silva_2019, LeBreton_2018, Kim_2022}.

    \begin{figure}
        \centering
        \includegraphics[scale=0.7]{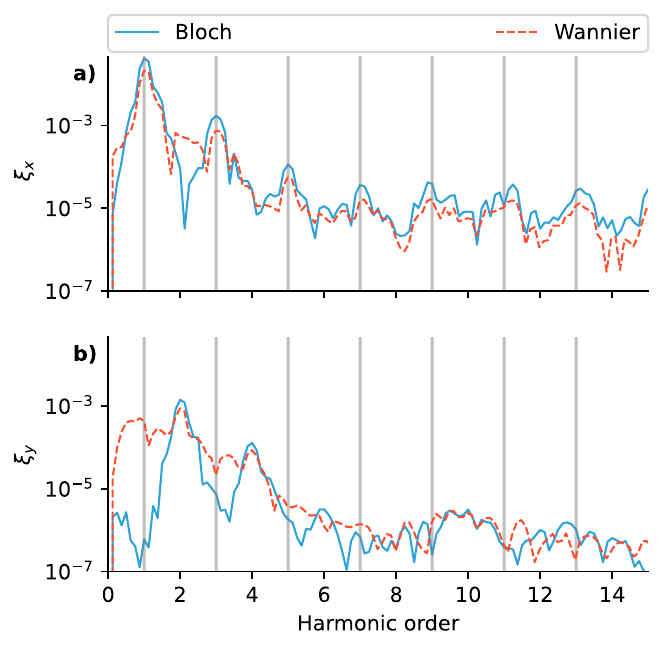}
        \caption[]{\Gls{hhg} spectra of the \gls{hbn} system along the $x$ and $y$ axis derived from the current
        calculated directly from the Bloch states or from the Wannier centers (\cref{eq:j_wan})}
        \label{fig:hbn_hhg}
    \end{figure}

    Looking at the \gls{hhg} spectra of this system in \cref{fig:hbn_hhg} we find a weaker but still relatively good
    agreement between the Wannier function derived spectra and the reference one along the driving axis $x$.
    The fact that the agreement is not as good as in the previous model case is to be expected as we are performing
    various conceptual leaps, in that we are now applying \gls{td-dft} functionals approximations, we are using a
    coarser time, space and $k$-space grid, we are evaluating a shorter laser driving, and we are evaluating the
    time-dependent Wannier functions less frequently.
    It is natural that more numerical errors accumulate, and in light of those factors, the fact that the \gls{hhg}
    spectra still agree that much, however, is promising.

    \begin{figure*}
        \centering
        \includegraphics[scale=0.7]{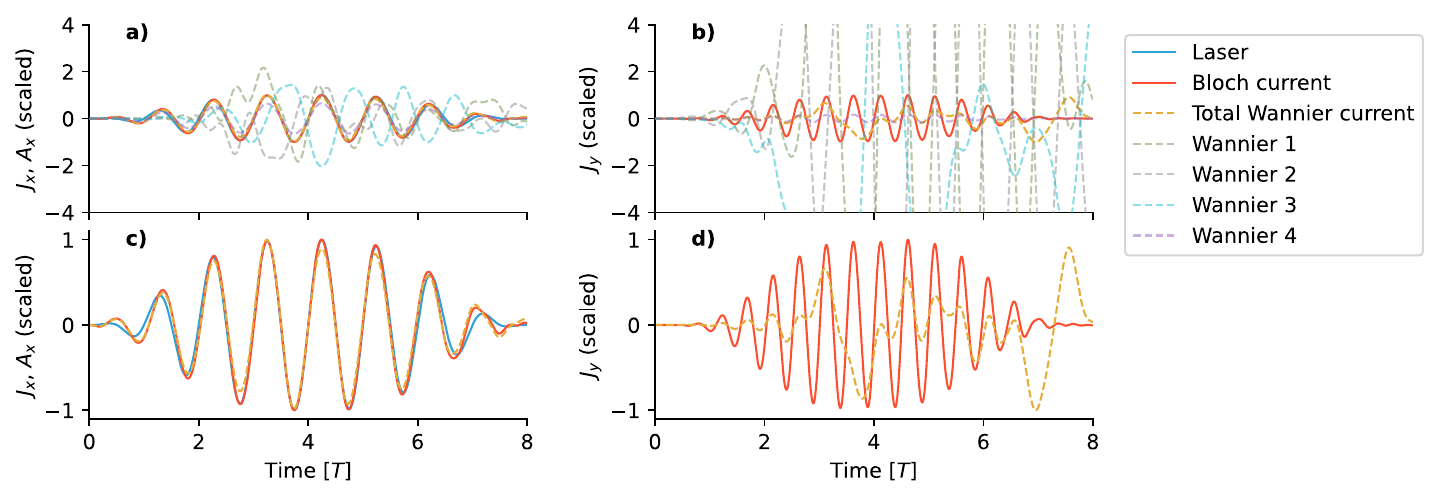}
        \caption[]{Applied laser and currents of the \gls{hbn} system calculated directly from the Bloch states or from
        the Wannier centers (\cref{eq:j_wan}).}
        \label{fig:hbn_t}
    \end{figure*}

    We next consider the \gls{hhg} spectra along the $y$-axis, where we see a stronger disagreement.
    To understand the cause of this deviation, we look at the raw time-dependent currents which we show in
    \cref{fig:hbn_t} along with the driving laser profile.
    Here we do not apply the Gaussian filtering, and we scale the laser driving, the reference current and the total
    Wannier current to their highest absolute value for plotting visibility.
    Here we want to point out how each Wannier function contribute to the total Wannier current.
    Along both the $x$ and $y$ axis we have significant cancellation of currents as we have similarly seen in the
    non-interacting model in \cref{fig:mathieu_hhg_wfs}, but in this case, the cancelling oscillations exceed the total
    current, especially along the $y$-axis where the difference is more than one order of magnitude.
    This, combined this with how the reference \gls{hhg} spectra along the $y$-axis is two orders of magnitude smaller,
    and the limitations of the simulations that we previously mentioned, it is acceptable to find such
    deviations, due to the numerical limitations.

    We should also disregard completely the theoretical limitations of the time-dependent Wannier functions defined from
    \cref{eq:td_wannier,eq:f_int}.
    Throughout this analysis, we have evaluated the Wannier current by considering the Wannier function of a single unit
    cell, effectively disregarding the contribution of the dynamics in the many-body representation.
    In \cref{sec:non-interacting-propagator} we evaluate the propagator of the non-interacting system to find one source
    of deviation due to resonant excitation.
    We expect that with further refinement of the time-dependent Wannier function definition and more converged
    simulation parameters, we would find a more optimal single-particle basis set to represent the dynamics of the
    many-body interacting system.

    \section{Conclusion}

    The main result of this work is the definition of the time-dependent Wannier functions that can be efficiently
    performed on top of well-established first-principles calculation methods, primarily the real-time \gls{td-dft}.
    In comparison with other proposed definitions for the time-dependent Wannier functions, the method developed here
    avoids the most expensive part of maximally localizing the basis set.
    The dynamical polarization model is shown to give good agreements of the current density calculations if we
    compensate for the dephasing factor of the Wannier states.

    While other time-dependent lattice basis, such as the Peierls basis set, require an extensive basis including the
    initial unoccupied bands and various lattice sites\cite{Ikemachi_2017, Liu_2017, Thorpe_2023}, with the
    time-dependent Wannier functions we are able to limit the necessary basis set to a few of the occupied bands and a
    single lattice site.
    This results in a much simpler lattice model that can also be used to study strongly correlated materials, besides
    the applications we demonstrate here.

    While we here have only explored the applicability of the time-dependent Wannier functions for extracting some
    physical intuition of the underlying dynamics and as a tool for analyzing non-linear electronic dynamics, there is a
    vast range of further applications to explore, from improving the \gls{td-dft} calculations, to visualizing and more
    efficiently calculating the dynamic observables.
    Looking forward, we believe that time-dependent Wannier functions will prove useful in the evolving field of
    ultrafast dynamics in solids, especially moving towards strongly correlated materials\cite{Silva_2018,
    TancogneDejean_2018, Valmispild_2024, Uchida_2022, Murakami_2022}, phonon dynamics\cite{Zhang_2024b, Bionta_2021,
    Neufeld_2022b, Lively_2024}, excitonic effects\cite{Kobayashi_2023, ChangLee_2024, Jensen_2024, Molinero_2024,
    Udono_2022, Freudenstein_2022, Cohen_2024}, superconductivity\cite{Alcal_2022, Hu_2024}, etc.
    These time-dependent basis can prove useful in state of the art calculation methods for highly correlated
    systems\cite{Seth_2016, Tai_2023}.

    \section{Acknowledgements}
    This work was supported by the Cluster of Excellence \enquote{CUI:Advanced Imaging of Matter} of the
    \gls{dfg} (EXC 2056 and SFB925), and the Max Planck-New York City Center for Non-Equilibrium Quantum Phenomena.
    The Flatiron Institute is a division of the Simons Foundation.

    \appendix

    \section{Relation to other basis sets}\label{sec:relation-to-other-basis}

    In order to examine the relation of the time-dependent Wannier functions with other commonly used basis like the
    Peierls basis or the Houston basis, we look at the time-dependent non-interacting electron system, for which we can
    express the time-dependent Schrödinger equation as
    \begin{gather}
        \psi^{\vb{A}}_{n\vb{k}}(\vb{r},t) = \ip{\vb{r}}{\psi^{\vb{A}}_{n\vb{k}}(t)} = e^{-i\frac{1}{\hbar}E^{0}_{n\vb{k}}t} e^{i(\vb{k}+\frac{e}{\hbar}\vb{A}(t))\vb{r}} u_{n\vb{k}}(\vb{r}), \label{eq:td_basis}\\
        C^{i}_{n\vb{k}}(t) = \ip{\psi^{\vb{A}}_{n\vb{k}}(t)}{\Psi_{i\vb{k}}(t)}, \\
        \partial_t C^{i}_{n\vb{k}}(t) = -e\pdv{\vb{A}(t)}{t}\sum_m\vb{D}_{nm}^{\vb{k}}(t) C^{i}_{m\vb{k}}(t), \label{eq:dt_C}
    \end{gather}
    where $\psi$ is an auxiliary basis set which we use here for decomposing the dynamics of the propagator,
    $C^{i}_{n\vb{k}}$ is the time-dependent projection of the Bloch wave function $\Psi_{i\va{k}}$ on the time-dependent
    wave function $\psi_{n\va{k}}$, and $\vb{D}_{nm}^{\vb{k}}(t)$ is constructed from the transition dipole moment
    $\vb{D}_{nm}^{\vb{k}}$
    \begin{gather}
        \vb{D}_{nm}^{\vb{k}}(t) = e^{-i\frac{1}{\hbar}(E^{0}_{m\vb{k}}-E^{0}_{n\vb{k}})t}\vb{D}_{nm}^{\vb{k}}, \label{eq:D_nm_t}\\
        \vb{D}_{nm}^{\vb{k}} = \int\dd{\vb{r}}u^*_{n\vb{k}}(\vb{r})\vb{r}u_{m\vb{k}}(\vb{r}).
    \end{gather}
    More details on this derivation are provided in \cref{sec:non-interacting-propagator}.
    In this derivation we are assuming the dipole approximation.
    Here we point out that the outside the resonance regime, \cref{eq:D_nm_t} oscillates rapidly such that
    \cref{eq:dt_C} approximates to a constant, and we can take \cref{eq:td_basis} to approximate the time-dependent
    Bloch states.

    We apply the basis \cref{eq:td_basis} to \cref{eq:td_wannier}
    \begin{align}
        \Phi_{n\vb{R}}(\vb{r},t) &\approx \sum_{i}\int_{BZ}U^{n\vb{R}}_{i\vb{k}}f_{i\vb{k}}(t)\psi^{\vb{A}}_{i\vb{k}}(\vb{r},t)\dd[3]{\vb{k}}, \\
        &\approx P^{\vb{A}C}(\vb{r},t)\sum_{i}\int_{BZ}P^{nD}_{i\vb{k}}(t)U^{n\vb{R}}_{i\vb{k}}\Psi'_{i\vb{k}}(\vb{r},t)\dd[3]{\vb{k}},
    \end{align}
    where we group the time-dependent effects into a phase acceleration $P^{\vb{A}C}$, common for all Wannier functions
    and dependent only on the laser profile, the main dephasing contribution $P^{D}$ that we investigated in
    \cref{sec:td-wannier-func}, and the remaining Bloch state dynamics $\Psi'$ that are covered in \cref{eq:dt_C}
    \begin{align}
        P^{\vb{A}C}(\vb{r},t) &= e^{i\frac{e}{\hbar}\vb{A}(t)\vb{r}},\\
        P^{nD}_{i\vb{k}}(t) &\approx e^{-i\frac{1}{\hbar}E^{0}_{n\vb{k}}t}f_{i\vb{k}}(t), \\
        \Psi'_{i\vb{k}}(\vb{r},t) &\approx \sum_{j} C^{j}_{i\vb{k}}(t) e^{i\vb{k}\vb{r}} u_{j\vb{k}}(\vb{r}), \\
    \end{align}
    $P^{\vb{A}C}$ does not contribute to the dynamics of Wannier centers or Wannier spread, $P^{D}$ is close to unity
    in the Wannier functions $f^{\text{int}}$, and the only remaining effects come from the excitation dynamics in
    \cref{eq:dt_C} through $\Psi'$.
    The dynamics in $\Psi'$ are the main dynamics that we see in \cref{sec:td-wannier-func,sec:results}.

    Comparing this expression with Peierls basis $\Phi^{P}$ and the Houston basis $\Psi^{H}$
    \begin{gather}
        \Phi^{P}_{n\vb{R}}(\vb{r}, t) = e^{i\frac{e}{\hbar}\vb{A}(t)\vb{r}}\Phi_{n\vb{R}}(\vb{r}, 0), \\
        \Psi^{H}_{i\vb{k}}(\vb{r}, t) = e^{i\frac{e}{\hbar}\vb{A}(t)\vb{r}}\Psi_{i\vb{k}+\vb{A}}(\vb{r}, 0),
    \end{gather}
    we can see how all of these basis sets merely account for the $P^{\vb{A}C}$ contribution, while the remaining
    dynamics have to be included by expanding the basis set into the unoccupied states and/or the many-body
    coefficients.
    When we consider that the Peierls substitution is often performed on a single band basis for computational
    efficiency, we can see how the dynamics in \cref{eq:dt_C} can be completely overlooked.

    \section{Non-interacting propagator}\label{sec:non-interacting-propagator}

    In \cref{sec:relation-to-other-basis} we have referenced the exact propagator of non-interacting systems, and in
    this appendix we will elaborate on how we have derived it for the sake of completeness.
    We start from the time-independent Hamiltonian
    \begin{equation}
        \hat{H}^0 = \frac{1}{2m}\hat{\vb{p}}^2 + \hat{V},
    \end{equation}
    where $\vb{p}$ and $V$ are the momentum and spatially periodic potential, respectively.
    Let's consider we have solved the time-independent Schrödinger equation to get the eigenstates and eigenvalues
    \begin{gather}
        \hat{H}^0 \ket{\Psi^0_{i\vb{k}}} = E^0_{i\vb{k}} \ket{\Psi^0_{i\vb{k}}},\\
        \Psi^0_{i\vb{k}}(\vb{r}) = \ip{\vb{r}}{\Psi^0_{i\vb{k}}} = e^{i\vb{k}\vb{r}} u_{i\vb{k}}(\vb{r}).
    \end{gather}

    For the laser interaction, we take the velocity gauge Hamiltonian
    \begin{equation}
        \hat{H}(t) = \frac{1}{2m}(\hat{\vb{p}} - e\vb{A}(t))^2 + \hat{V},
    \end{equation}
    where $e$ is the charge of the electron and $\vb{A}$ is the applied laser field.
    For simplicity, we are taking the dipole approximation, ignoring the spatial variation of $\vb{A}$.
    We solve the time-dependent Schrödinger equation using the basis defined in \cref{eq:td_basis}.
    This basis is an instantaneous eigenbasis of the time-dependent Hamiltonian
    \begin{equation}
        \mel{\psi^{\vb{A}}_{m\vb{k}'}(t)}{\hat{H}(t)}{\psi^{\vb{A}}_{n\vb{k}}(t)} = \delta_{mn}\delta(\vb{k}'-\vb{k})E^0_{n\vb{k}}.
    \end{equation}
    We solve the time-dependent Schrödinger equation using the overlap
    \begin{equation}
        C^{i\vb{k}}_{n\vb{k}'}(t) = \ip{\psi^{\vb{A}}_{n\vb{k}'}(t)}{\Psi_{i\vb{k}}(t)},
    \end{equation}
    where $\Psi_{i\vb{k}}(t)$ is the time-dependent wave function of the non-interacting electron with the initial
    condition
    \begin{equation}
        \ket{\Psi_{i\vb{k}}(0)} = \ket{\Psi^0_{i\vb{k}}}.
    \end{equation}

    Expanding the time-dependent Schrödinger equation on this basis we get
    \begin{widetext}
        \begin{gather}
            \mel{\psi^{\vb{A}}_{n\vb{k}'}(t)}{i\hbar\partial_t}{\Psi_{i\vb{k}}(t)} = \mel{\psi^{\vb{A}}_{n\vb{k}'}(t)}{\hat{H}(t)}{\Psi_{i\vb{k}}(t)}, \\
            \sum_m\int\dd{\vb{k}''} \mel{\psi^{\vb{A}}_{n\vb{k}'}(t)}{i\hbar\partial_t}{\psi^{\vb{A}}_{m\vb{k}''}(t)} \ip{\psi^{\vb{A}}_{m\vb{k}''}(t)}{\Psi_{i\vb{k}}(t)} = E^0_{n\vb{k}'} C^{i\vb{k}}_{n\vb{k}'}(t), \\
            \sum_m\int\dd{\vb{k}''} \qty\Big[-e\pdv{\vb{A}(t)}{t}\int\dd{\vb{r}}\mel{\psi^{\vb{A}}_{n\vb{k}'}(t)}{\hat{\vb{r}}}{\psi^{\vb{A}}_{m\vb{k}''}(t)} + \delta_{mn}\delta(\vb{k}'-\vb{k}'')\qty(i\hbar\partial_t + E^0_{m\vb{k}''})] C^{i\vb{k}}_{m\vb{k}''}(t) = E^0_{n\vb{k}'} C^{i\vb{k}}_{n\vb{k}'}(t). \label{eq:td-shcro3}
        \end{gather}
    \end{widetext}
    We denote $\vb{D}_{nm}^{\vb{k}'\vb{k}''}(t)$ constructed from the transition dipole moment
    $\vb{D}_{nm}^{\vb{k}'\vb{k}''}$ as
    \begin{align}
        \vb{D}_{nm}^{\vb{k}'\vb{k}''}(t) &= \mel{\psi^{\vb{A}}_{n\vb{k}'}(t)}{\hat{\vb{r}}}{\psi^{\vb{A}}_{m\vb{k}''}(t)}, \\
        &= e^{-i\frac{1}{\hbar}(E^{0}_{m\vb{k}''}-E^{0}_{n\vb{k}'})t}\vb{D}_{nm}^{\vb{k}'\vb{k}''}, \\
        \vb{D}_{nm}^{\vb{k}'\vb{k}''} &= \delta(\vb{k}'-\vb{k}'')\int\dd{\vb{r}}u^*_{n\vb{k}}(\vb{r})\vb{r}u_{m\vb{k}}(\vb{r}).
    \end{align}
    Substituting this into \cref{eq:td-shcro3} and simplifying the dirac deltas we get
    \begin{equation}
        \partial_t C^{i\vb{k}}_{n\vb{k}'}(t) = -e\pdv{\vb{A}(t)}{t}\sum_m\vb{D}_{nm}^{\vb{k}'}(t) C^{i\vb{k}}_{m\vb{k}'}(t), \label{eq:dt_Cb}
    \end{equation}
    or considering the initial condition
    \begin{equation}
        C^{i\vb{k}}_{n\vb{k}'}(0) = \delta_{in}\delta(\vb{k}-\vb{k}'),
    \end{equation}
    we can simplify the equation of motion by ignoring $\vb{k}'\neq\vb{k}$ and we get the form in \cref{eq:dt_C}.

    \bibliography{TDWannier}
\end{document}